\documentclass{article}

\usepackage{arxiv}

\usepackage[utf8]{inputenc} 
\usepackage[T1]{fontenc}    
\usepackage{hyperref}       
\usepackage{url}            
\usepackage{booktabs}
\usepackage{array}       
\usepackage{amsfonts}       
\usepackage{microtype}      
\usepackage{lipsum}
\usepackage{graphicx}
\usepackage{pgfplots}
\usetikzlibrary{arrows.meta,shapes.geometric}
\pgfplotsset{compat=1.12}
\graphicspath{ {./images/} }

\renewcommand{\today}{November 18, 2026}
\title{A Brief Overview of Non-Terrestrial Networks (Updated 2026)}
\hypersetup{
  pdftitle={A Brief Overview of Non-Terrestrial Networks (Updated 2026)},
  pdfauthor={Jos\'{e} Alberto Hern\'{a}ndez and Pedro Reviriego}
}

\author{
  Jos\'{e} Alberto Hern\'{a}ndez\\
  Universidad Carlos III de Madrid\\
  28911 Madrid, Spain \\
  \texttt{jahgutie@it.uc3m.es} \\
  \And  
  Pedro Reviriego \\
  Universidad Polit\'{e}cnica de Madrid\\
  28040 Madrid, Spain \\
  \texttt{pedro.reviriego@upm.es} \\
}

\begin{document}

\maketitle

\begin{abstract}
This article extends the 2023 preprint \emph{A brief introduction to satellite communications for Non-Terrestrial Networks (NTN)} (\href{https://arxiv.org/abs/2305.04590v1}{arXiv:2305.04590v1}) with a revision of its foundations, tables and state of the art as of 18 November 2026. It revisits propagation, link budgets, spectrum reuse and phased arrays, including corrected equations and numerical examples. Updated tables distinguish operational, launched and planned constellations and compare published fixed-terminal rates with direct-to-device services. The discussion incorporates developments in 3GPP Releases 17--20, commercial LTE-based direct-to-cell services and research towards 6G non-terrestrial networks. Operator announcements are distinguished from independent measurements and future plans. Open challenges include spectrum coordination, Doppler and timing compensation, mobility, capacity, security and the sustainability of low Earth orbit.
\end{abstract}



\section{Introduction}
\label{sec:introduction}

This article is a continuation and substantial update of our 2023 preprint, \emph{A brief introduction to satellite communications for Non-Terrestrial Networks (NTN)}~\cite{original2023}. It retains the tutorial treatment of satellite communications while revising its equations and numerical examples, updating tables on constellation deployment and service capabilities, and incorporating developments in direct-to-device connectivity and the integration of NTNs with 5G and emerging 6G systems. The deployment figures and service descriptions in this update reflect the cited sources available as of 18 November 2026; announced plans and advertised performance are identified as such.

Terrestrial fibre deployment can be expensive where users are sparse, particularly when civil works dominate costs; infrastructure sharing changes the investment trade-offs~\cite{schneir_2014}. Satellite networks can complement terrestrial access in remote, maritime and mountainous areas. Low Earth orbit (LEO) constellations are a prominent example.

The scale of LEO broadband adoption has grown rapidly. SpaceX reported 13.6 million Starlink consumer subscribers at the end of September 2026, compared with 7.2 million a year earlier, an increase of approximately 89\%~\cite{spacex_10q2026}. The metric counts unique service lines and includes some business plans, while excluding managed enterprise and government contracts; it does not count individual people. By comparison, ITU estimated approximately 1.6 billion fixed-broadband subscriptions worldwide in 2025~\cite{itu_broadband2025}. The dates and subscription definitions differ, and the ITU fixed-broadband category can include satellite access~\cite{itu_ict_handbook}; these figures should not be interpreted as a measure of satellite traffic share.

Measures of traffic and capacity point to growing demand but must be interpreted at their measurement points. In its 2025 review, Cloudflare reported a 2.3-fold increase during 2025 in request volume observed from Starlink's principal autonomous system, AS14593~\cite{cloudflare2025}. This is a provider-specific count of requests to Cloudflare, not the total number of bytes carried across Starlink. For historical scale, a 2024 SpaceX technical presentation described optical inter-satellite terminals operating at 100 Gb/s per link~\cite{brashears2024}; an April 2025 MIT Lincoln Laboratory presentation reported more than 42 PB/day traversing Starlink laser links, referencing that earlier presentation~\cite{robinson2025}. The 42 PB/day figure is a historical inter-satellite traffic measure, not a 2026 measurement of unique end-user traffic, and traffic crossing several links may be counted differently from access-network traffic. Likewise, a link's maximum bit rate is a capacity figure, not its average traffic or a rate guaranteed to each subscriber. SpaceX separately reported a 225 Mb/s median peak-hour residential download speed as of March 2026; this is an operator-reported user-performance statistic, not aggregate system capacity~\cite{spacex_prospectus2026}.

For context, ITU estimates that in 2026 worldwide end-user traffic reached 7.5 ZB on fixed broadband and 1.7 ZB on mobile broadband, with fixed traffic rising from a revised estimate of 6.6 ZB in 2025. Since 2021, fixed and mobile broadband traffic grew at average annual rates of 13.6\% and 18.2\%, respectively~\cite{itu_traffic2026}. The 2026 figures are estimates based on extrapolation of country-level trends~\cite{itu_method2026}. These statistics are grouped by access category, not by a mutually exclusive satellite-versus-terrestrial classification: fixed broadband can include satellite access, and Wi-Fi traffic from phones is counted with the fixed connection. Consequently, neither ITU's aggregates nor Cloudflare's request counts provide a directly comparable global byte share for LEO networks. Together, subscriber growth and observed request growth indicate expanding use, but they do not establish the same percentage growth in total satellite-carried bytes.


Falling launch costs and mass production have accelerated LEO deployments, although economics depend on launch, replacement and ground infrastructure~\cite{techno_economic_sats}. In October 2026 SpaceX reported more than 11,000 Starlink satellites in orbit; this figure is operator-specific and is not a count of all active satellites~\cite{starlink_update}.

Round-trip time (RTT) depends on orbit altitude and the end-to-end route. For a transparent geostationary Earth orbit (GEO) relay, four space--ground traversals of at least 35,786 km give a propagation-only lower bound of about 477 ms for the user-to-gateway round trip. Medium Earth orbit (MEO) reduces this distance, and LEO reduces it further. A high-altitude platform (HAP) at about 20 km can shorten the propagation path substantially. Ground routing, scheduling and queueing add to these geometric delays~\cite{tr38811}.

Signals propagate at about 300,000 km/s in free space and about 200,000 km/s in typical silica fibre, corresponding to roughly $3.33~\mu\mathrm{s/km}$ and $5~\mu\mathrm{s/km}$. LEO paths can have lower propagation delay than fibre over some long routes. The crossover distance depends on satellite and gateway geometry, inter-satellite routing and the length of the terrestrial fibre route~\cite{Handley_2018,Handley_2019}.

Several broadband constellations use orbits between about 500 and 1,200 km to reduce propagation delay~\cite{tr38811,oneweb2026}. In addition to providing coverage to rural areas, satellite networks can complement terrestrial coverage during disasters, subject to power, gateway, regulatory and security constraints. LEO satellite constellations can provide sufficient connectivity performance for Machine-Type Communications (MTC) and Mobile Broadband (MBB) in such remote areas~\cite{ngmn_ntn}, paving the way for Non-Terrestrial Networks (NTN) that complement existing Terrestrial Networks, both fixed and mobile. The authors of~\cite{guidotti_LTE,guidotti_5g} provide a summary of architectures and challenges to integrate LEO constellations in the 5G ecosystem and even 6G~\cite{aranity_2021}. Rinaldi et al. survey the architecture and service scenarios of NTN in 5G and beyond~\cite{ieeeaccess_survey}.

SpaceX/Starlink and Eutelsat OneWeb operate large LEO broadband systems; Amazon Leo (formerly Project Kuiper) is deploying its network, while Telesat Lightspeed is at a pre-service stage~\cite{starlink_update,oneweb2026,amazon2026,telesat2026}. Earlier comparisons such as~\cite{DELPORTILLO2019123,hypatia} remain useful for methodology but their constellation designs are historical.

This article provides a brief review on the basic principles and design requirements of satellite communications. This review contains several numerical examples to give the reader a better feeling of the uses and applications of satcoms. To this end, Section~\ref{sec:soa} briefly reviews the most important design aspects of Satellite Communications. Section~\ref{sec:stin} introduces current mega-constellations and existing projects for NTNs as of November 2026. 
Finally, Section~\ref{sec:conclusions} concludes this work with a summary of its main contributions.


\section{An overview of satellite communications}
\label{sec:soa}

\subsection{Orbits and propagation delay}

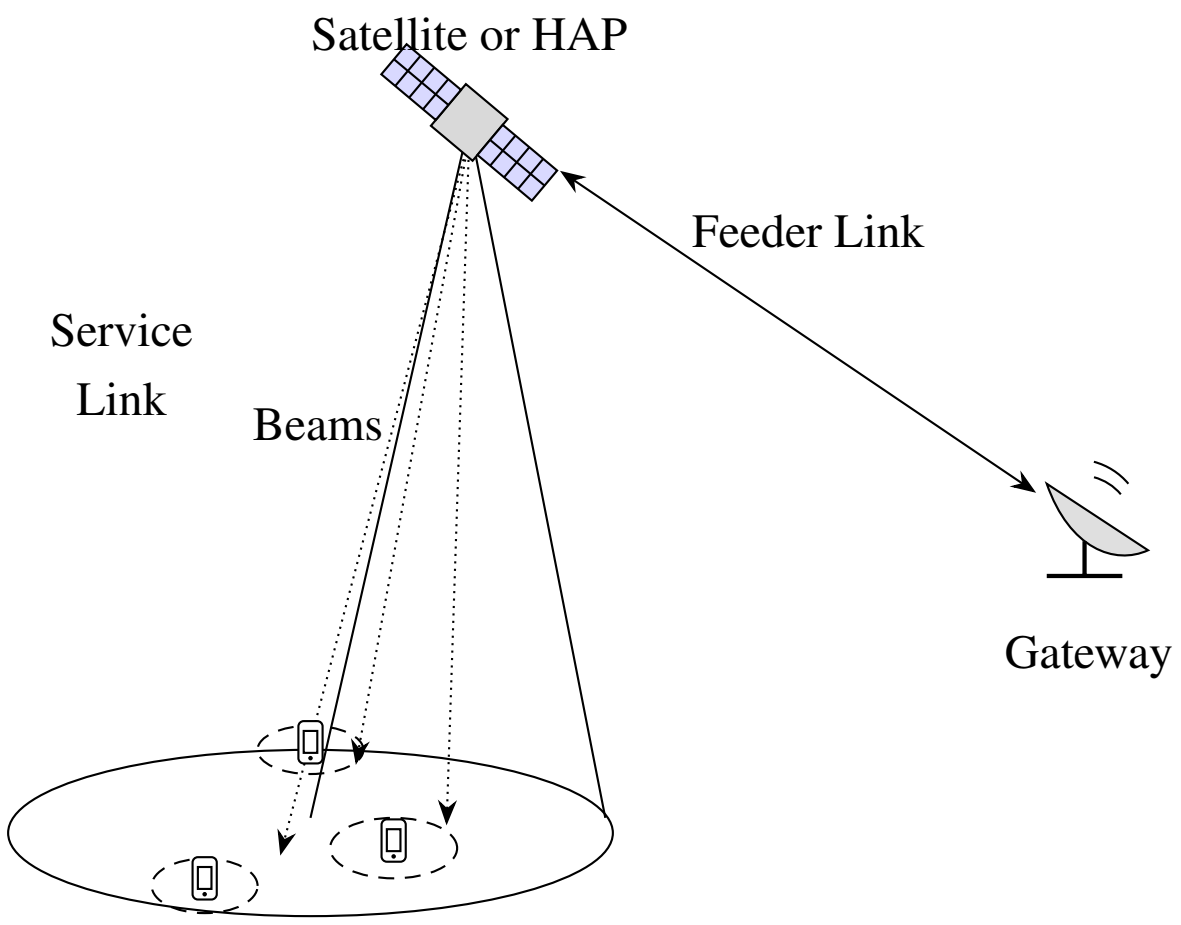
\begin{figure}[htbp]
\centering
\begin{tikzpicture}[x=1cm,y=1cm,>=Stealth, font=\small]
\draw (0,0) ellipse (2.0 and 0.55);
\draw (0,0.1) -- (1.05,4.7) -- (1.95,0.1);
\draw[densely dotted,->] (1.05,4.6) -- (0.3,0.45);
\draw[densely dotted,->] (1.05,4.6) -- (0.9,0.05);
\draw[densely dotted,->] (1.05,4.6) -- (-0.2,-0.15);
\draw[densely dashed] (-0.7,-0.35) ellipse (0.35 and 0.18);
\draw[densely dashed] (0.55,-0.1) ellipse (0.42 and 0.2);
\draw[densely dashed] (0,0.55) ellipse (0.35 and 0.16);
\foreach \x/\y in {-0.7/-0.30,0.55/-0.05,0/0.60}{
  \draw[rounded corners=1pt,fill=white] (\x-0.08,\y-0.14) rectangle (\x+0.08,\y+0.14);
  \draw (\x-0.045,\y-0.07) rectangle (\x+0.045,\y+0.075);
  \fill (\x,\y-0.105) circle (0.015);
}
\begin{scope}[shift={(1.05,4.7)},rotate=-40]
  \draw[fill=gray!30] (-0.18,-0.18) rectangle (0.18,0.18);
  \draw[fill=blue!15] (-0.65,-0.13) rectangle (-0.18,0.13);
  \draw[fill=blue!15] (0.18,-0.13) rectangle (0.65,0.13);
  \foreach \x in {-0.53,-0.41,-0.29,0.29,0.41,0.53} {\draw (\x,-0.13)--(\x,0.13);}
  \draw (-0.65,0)--(-0.18,0) (0.18,0)--(0.65,0);
\end{scope}
\node[above] at (1.05,5.05) {Satellite or HAP};
\draw[<->] (1.65,4.38) -- (4.8,2.25);
\node[above right] at (2.4,3.75) {Feeder Link};
\begin{scope}[shift={(5.12,1.85)}]
  \draw[fill=gray!25] (-0.25,0.46) .. controls (-0.1,0.03) and (0.20,-0.08) .. (0.42,0.02) -- cycle;
  \draw[thick] (0,0.08)--(0,-0.15) (-0.25,-0.15)--(0.25,-0.15);
  \draw (0.24,0.39) arc (40:75:0.35) (0.29,0.46) arc (40:75:0.45);
\end{scope}
\node[below] at (5.15,1.43) {Gateway};
\node[align=center] at (-1.25,3.1) {Service\\Link};
\node at (0.05,2.7) {Beams};
\end{tikzpicture}
\caption{Architecture and terminology.} 
\label{fig:arch}
\end{figure}

In general Non-Terrestrial Networks (NTNs) refer to networks providing connectivity through space-borne vehicles or airborne platforms, including satellites, High-Altitude Platforms (HAPs) and Low-Altitude Platforms (LAPs). These provide radio connectivity between the User Equipment (UE) on the ground and the vehicle which, in addition, provide connectivity to Terrestrial Networks (TN) through Ground Based Gateways (see Fig.~\ref{fig:arch}).

Depending on the altitude of the space-borne, multiple NTN options are possible:
\begin{itemize}
    \item Stationary satellites placed in GEO, operating at 35,786~km altitude. GEO scenarios are often equipped with Very High Throughput Satellites (VHTS), providing tens or hundreds of Gb/s capacity each. In this case, Doppler effects are negligible but propagation delays can reach up to several hundreds of millisecs for transparent satellites. 
    \item LEO extends to about 2,000 km, and MEO lies between LEO and GEO~\cite{nasa_orbits}. Satellites in these orbits move relative to users. Depending on beam steering, cells can be fixed on Earth for part of a pass or move across the surface. The network must account for Doppler and handovers as visibility and coverage change~\cite{tr38811}.
    \item High Altitude Platforms (HAPs) like planes or balloons, operating like satellites but closer to the Earth, at about 20~km distance. The short radio path reduces propagation delay.
    \item Low Altitude Platforms (LAPs) like drones or balloons at less than 1~km altitude.
\end{itemize}

The altitude of the space-borne has a clear impact on the round-trip time. In this regard, it is worth remarking that latency heavily affects TCP throughput in TCP/IP based networks, which the Mathis model approximates for congestion avoidance under light to moderate loss~\cite{mathis}:
\begin{equation}
    \label{eq:mathis}
    Throughput_{TCP} \approx \frac{MSS_{bits}}{RTT} \frac{C}{\sqrt{p_{loss}}}
\end{equation}
where $MSS_{bits}$ is the maximum segment size in bits, RTT is the end-to-end round-trip time, $C$ is a model-dependent constant and $p_{loss}$ is the rate of loss indications per packet. With independent, isolated losses it can be approximated by the packet-loss probability. This is an approximate steady-state model for a long-lived congestion-avoidance TCP flow with sufficient receiver window and few retransmission timeouts, not a universal upper bound: receiver windows, link capacity, congestion-control algorithm, timeouts and loss correlations can change the result~\cite{mathis,padhye_tcp}.

\noindent\fbox{%
\parbox{\dimexpr\linewidth-2\fboxsep-2\fboxrule-3pt\relax}{
\textbf{Numerical example no. 1:} Assume RTT $=0.2$ s, $MSS=1500$ bytes $=12000$ bits, $C=1$ and $p_{loss}=10^{-9}$. The model gives approximately 1.897 Gb/s. For RTT $=0.4$ s with the same loss, it gives 0.949 Gb/s; for RTT $=0.4$ s and $p_{loss}=10^{-6}$, it gives 30 Mb/s. These figures illustrate the model and are not guaranteed measured throughputs.
}%
}

\subsection{Frequency bands}

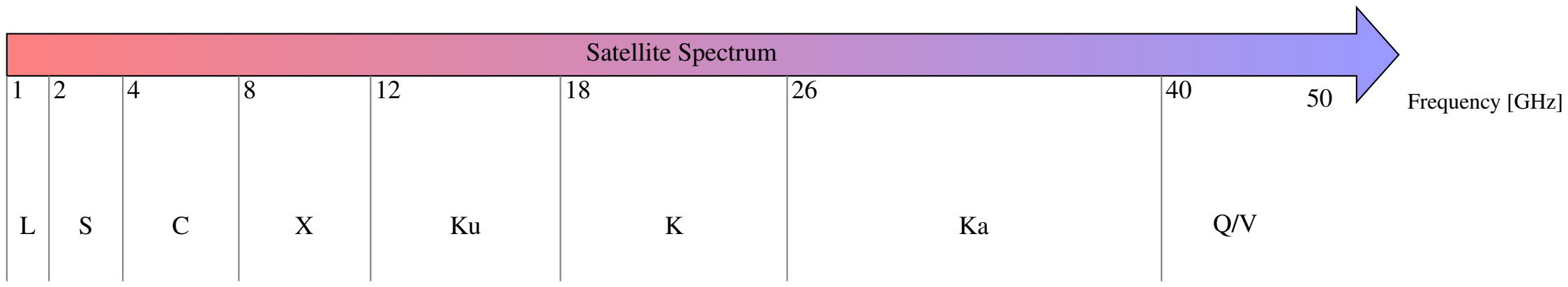
\begin{figure}[htbp]
\centering
\begin{tikzpicture}[x=1cm,y=1cm,font=\scriptsize]
\shade[left color=red!50,middle color=green!35,right color=blue!40] (0,1) -- (12.8,1) -- (12.8,1.25) -- (13.2,0.8) -- (12.8,0.35) -- (12.8,0.6) -- (0,0.6) -- cycle;
\draw (0,1) -- (12.8,1) -- (12.8,1.25) -- (13.2,0.8) -- (12.8,0.35) -- (12.8,0.6) -- (0,0.6) -- cycle;
\node at (6.4,0.8) {Satellite Spectrum};
\foreach \x/\v in {0/1,0.4/2,1.1/4,2.2/8,3.45/12,5.25/18,7.4/26,10.95/40}{\draw[gray] (\x,0.6)--(\x,-1.35); \node[anchor=north west,inner sep=1pt] at (\x,0.6) {\v};}
\node[anchor=north] at (12.45,0.6) {50};
\foreach \x/\name in {0.2/L,0.75/S,1.65/C,2.82/X,4.35/Ku,6.33/K,9.17/Ka} {\node at (\x,-0.82) {\name};}
\node at (11.65,-0.82) {Q/V};
\node[anchor=west,font=\tiny] at (13.16,0.34) {Frequency [GHz]};
\end{tikzpicture}
\caption{Conventional band labels in the original schematic~\cite{IEEE_surv_tutorial_kodheli}. The horizontal scale is schematic. Table~\ref{tab:freqbands} gives selected operating ranges and their link directions.} 
\label{fig:bands}
\end{figure}

Figure~\ref{fig:bands} illustrates frequencies up to 50 GHz; Table~\ref{tab:freqbands} adds higher-frequency examples and identifies the link direction and intended use. In this table downlink (DL) is space-to-Earth and uplink (UL) is Earth-to-space. The entries are selected examples, not a comprehensive allocation table. Frequency rights depend on the service, orbit, jurisdiction, coordination and licence; the listed endpoints must not be treated as universal permission to operate.

For 5G NR-NTN, 3GPP specifies bands in L, S, Ku and Ka, including n255 in L, n256 in S and n254 with an L-band uplink and S-band downlink~\cite{etsi_ntn_bands}. The standard specifies equipment operating ranges but does not establish that every network or handset supports them. Direct-to-cell systems may also use terrestrial mobile spectrum under specific regulatory and operator arrangements; these are distinct from the NR-NTN bands. Ku and Ka provide broadband access and feeder links, while rain attenuation and coordination constrain availability. X-band is also used by certain LEO missions, particularly for telemetry and Earth-observation data~\cite{esa_xband_leo}. Optical inter-satellite links use light rather than the RF bands shown here~\cite{brashears2024}.

\begin{table}[htbp]
\centering\small
\caption{Selected satellite frequency examples. DL: space-to-Earth; UL: Earth-to-space. Values in GHz. Entries represent specified or authorised ranges, not proof of commercial deployment or global authorisation.}
\label{tab:freqbands}
\begin{tabular}{p{2.5cm}p{2.85cm}p{2.85cm}>{\raggedright\arraybackslash}p{5.1cm}}
\toprule
Band/example & DL (GHz) & UL (GHz) & Context and scope \\
\midrule
L / n255 & 1.525--1.559 & 1.6265--1.6605 & 3GPP NR-NTN operating range~\cite{etsi_ntn_bands}. \\
L/S / n254 & 2.4835--2.5000 & 1.6100--1.6265 & 3GPP NR-NTN cross-band pair~\cite{etsi_ntn_bands}. \\
S / n256 & 2.170--2.200 & 1.980--2.010 & 3GPP NR-NTN operating range~\cite{etsi_ntn_bands}. \\
PCS (US) & 1.990--1.995 & 1.910--1.915 & Example FCC-authorised Starlink direct-to-cell spectrum~\cite{fcc_gen22026}. \\
C (FSS) & 3.4--3.95; 4.5--4.8 & 5.725--6.275; 6.725--7.025 & Illustrative FSS ranges; allocations vary by region~\cite{itu_sat_handbook}. \\
X (FSS) & 7.25--7.75 & 7.925--8.425 & Governmental and military satellite links~\cite{itu_sat_handbook}. \\
X (LEO EO) & 8.025--8.400 & 7.190--7.250 & ESA example for LEO telemetry, commands and data; not a broadband access band~\cite{esa_xband_leo}. \\
Ku (FSS) & 10.7--12.75 & 12.75--13.25; 13.75--14.5 & Broadband examples; sub-band use and direction depend on service and region~\cite{itu_sat_handbook}. \\
Ku / n508, n509 & 10.70--12.75 & 13.75--14.00 (n508)\newline 14.00--14.50 (n509) & NR-NTN ranges with regional downlink limits~\cite{etsi_ntn_bands}. \\
Ka / n510 & 17.3--20.2 & 27.5--28.35 & Example 3GPP NR-NTN range with geographical conditions~\cite{etsi_ntn_bands}. \\
Ka (NGSO) & 17.8--18.6; 18.8--19.3 & 27.5--29.1; 29.5--30.0 & Example Starlink gateway authorisation in the US~\cite{fcc_gateway2026}. \\
Q/V (NGSO) & 37.5--42.0 & 47.2--50.2; 50.4--51.4 & Selected ranges in an FCC NGSO authorisation, subject to conditions~\cite{fcc_gen22026}. \\
E (NGSO) & 71--76 & 81--86 & Example FCC-authorised gateway bands; availability is licence-dependent~\cite{fcc_gateway2026,fcc_gen22026}. \\
W (NGSO) & -- & 92--94, 94.1--95\newline 95--100 & Selected conditional SpaceX uplink ranges, non-conforming and non-interference under ITU RR 4.4~\cite{fcc_gen22026}. \\
\bottomrule
\end{tabular}
\end{table}

The E-band gateway entries correspond to specific FCC authorisations, including a temporary 2026 earth-station grant; they do not by themselves demonstrate continuous operational use. The W-band entries are examples from an FCC authorisation on an unprotected, non-interference basis under Article 4.4 of the ITU Radio Regulations~\cite{fcc_gen22026}. Atmospheric attenuation, particularly rain fade, becomes more important at high frequencies; this is a limitation on Earth-space paths but does not apply in the same way to an inter-satellite path above the atmosphere.

\subsection{Link-budget calculations}


The Friis free-space transmission equation for matched, co-polarised antennas in the far field is~\cite{friis1946,itu_p525}:
\begin{equation}
\label{eq:friis}
P_r=P_t G_t G_r\left(\frac{\lambda}{4\pi d}\right)^2.
\end{equation}
Here $P_t$ and $P_r$ are powers in W at the reference planes, $G_t$ and $G_r$ are linear antenna gains in the relevant directions for the assumed matched ports, $d$ is slant range, $\lambda=c/f$, and $P_tG_t$ is EIRP. Atmospheric, polarisation and other losses are included separately. Gains and losses must refer to consistent planes.

For equivalent system noise temperature $T_{sys}$ and receiver noise-equivalent bandwidth $B_w$, the input-referred noise power is $N=k_B T_{sys}B_w$~\cite[eq.~(2.2-3), referred to the input]{nasa_noise}. Receiver noise figure $NF$ in dB determines the receiver's equivalent input noise temperature $T_e$~\cite[eqs.~(2.2-4) and (2.2-29)--(2.2-30)]{nasa_noise}:
\begin{equation}\label{eq:noise_temperature}
T_e=T_0\left(10^{NF/10}-1\right),\qquad T_{sys}=T_A+T_e,
\end{equation}
where $T_0=290$ K and $T_A$ is the antenna noise temperature referred to the same input plane. This simplified expression neglects feeder loss; if present, feeder attenuation and its physical temperature must also be included. The Boltzmann constant is $k_B=1.380649\times10^{-23}$ J/K, or $k_{dB}\simeq-228.60$ dBW/(K Hz). With $G/T=G_{r,dBi}-10\log_{10}(T_{sys}/1\,\mathrm K)$ and $B_{dBHz}=10\log_{10}(B_w/1\,\mathrm{Hz})$, the downlink SNR is~\cite{kodheli_linkbudget}\cite[\S6.1.3.1]{tr38821}:
\begin{equation}
\label{eq:linkbudget}
SNR_{dB}=EIRP_{dBW}+(G/T)_{dB/K}-L_{FS}-L_{atm}-L_{other}-B_{dBHz}-k_{dB}.
\end{equation}
This expression assumes that the stated EIRP is total power in the relevant signal bandwidth and that added losses are not already included in EIRP or $G/T$. For illustration, with $T_A=290$ K and no feeder loss, a receiver with $NF=7$ dB has $T_e=1163.4$ K and $T_{sys}=1453.4$ K; $G_r=0$ dBi gives $G/T=-31.62$ dB/K. For $NF=5$ dB and $G_r=12$ dBi, $T_{sys}=917.1$ K and $G/T=-17.62$ dB/K. A receiver with total $T_{sys}=290$ K and $G_r=0$ dBi has $G/T=-24.62$ dB/K.

The free-space loss in dB is~\cite{itu_p525}
\begin{equation}\label{eq:fspl}
L_{FS}=20\log_{10}\left(\frac{4\pi df}{c}\right).
\end{equation}
For a spherical Earth, ground-terminal elevation angle $\alpha$, satellite altitude $h$ and Earth radius $R_E$, the slant range is~\cite[eq.~(6.6-3)]{tr38811}:
\begin{equation}\label{eq:slant_range}
d=-R_E\sin\alpha+\sqrt{R_E^2\sin^2\alpha+h^2+2R_Eh}.
\end{equation}
The altitude itself equals slant range only for a satellite at zenith ($\alpha=90^\circ$); at lower elevation the longer path must be used. Atmospheric losses depend on frequency, geometry and weather; an availability assessment needs a propagation model rather than a fixed universal margin~\cite{crane_model}.

\noindent\fbox{%
\parbox{\dimexpr\linewidth-2\fboxsep-2\fboxrule-3pt\relax}{
\textbf{Numerical example no. 2:} Let $R_E=6371$ km, $h=21000$ km, $\alpha=1.2^\circ$, $f=2$ GHz ($\lambda\simeq0.15$ m), $P_t=26.6$ W, $G_t=13$ dBi and $G_r=0$ dBi. Equations~\ref{eq:slant_range} and~\ref{eq:friis} give $d\simeq26486$ km and a free-space received power of $1.08\times10^{-16}$ W. Using $c=299792458$ m/s, the EIRP is $27.25$ dBW and the free-space loss is $186.93$ dB. Assume $T_A=290$ K, $NF=7$ dB, no feeder loss, $B_w=1$ kHz and combined atmospheric/other losses of 9.6 dB. Then $G/T=-31.62$ dB/K and
\[SNR_{dB}\simeq27.25-31.62-186.93-9.6-30+228.60=-2.30\ \mathrm{dB}.\]
Before rounding, the SNR is $-2.3047$ dB, equivalent to a linear ratio of $0.5882$. The $1.2^\circ$ elevation illustrates the effect of slant range. Actual availability depends on terrain and the network elevation mask.
}%
}

\subsection{Bitrates, Shannon's capacity limit and Adaptive Coding and modulation}

The SNR from Eq.~\ref{eq:linkbudget} and the channel bandwidth determine the capacity $R_{max}$ of the ideal band-limited additive white Gaussian noise (AWGN) channel~\cite[Part IV, \S25, Theorem 17]{shannon1948}:
\begin{equation}
    R_{eff} \leq R_{max} = B_w \cdot \log_2(1+snr) = B_w \cdot \beta_{max}
    \label{eq:shannon}
\end{equation}
where the effective bitrate $R_{eff}$ used in transmission cannot be larger than the Shannon's limit $R_{max}$. 

Here $snr$ is the linear signal-to-noise power ratio. The maximum spectral efficiency $\beta_{max}$ is the channel capacity per unit bandwidth in bit/s/Hz:
\begin{equation}
    \label{eq:shannon_beta}
    \beta_{eff} \leq \beta_{max} = \log_2(1+snr)
\end{equation}

Operational spectral efficiency depends on waveform, link conditions, coding, interference and implementation losses; altitude alone does not determine it. 

\noindent\fbox{%
  \parbox{\dimexpr\linewidth-2\fboxsep-2\fboxrule-3pt\relax}{
    \textbf{Numerical example no. 3:} Using a channel bandwidth equal to the $1$ kHz noise-equivalent bandwidth in Example 2, Eq.~\ref{eq:shannon} gives $R_{max}=1000\log_2(1+0.5882)\simeq667$ bit/s under the AWGN model.
   }
}

Inverting Eq.~\ref{eq:shannon_beta} gives the minimum linear SNR for a target spectral efficiency:
\begin{equation}\label{eq:shannon_snr}
    snr_{eff} \geq snr_{req} = 2^{\beta_{eff}}-1
\end{equation}
A practical link budget uses the selected waveform's required SNR at the target error rate and an availability margin based on expected propagation losses. 

Under the ideal Shannon--Hartley model for an additive white Gaussian noise channel, a spectral efficiency $\beta$ requires $\mathrm{SNR}\geq 2^\beta-1$ in linear units. Table~\ref{tab:acm} gives theoretical lower bounds; practical modulation and coding schemes require additional margin and have implementation-specific thresholds.

\begin{table}[htbp]
\centering
\caption{Ideal Shannon-limit SNR for selected spectral efficiencies; practical modem thresholds are higher.}
\label{tab:acm}
\begin{tabular}{cc}
\toprule
Spectral efficiency (bit/s/Hz) & Minimum ideal SNR (dB) \\
\midrule
0.5 & $-3.83$ \\
1.0 & $0.00$ \\
1.5 & $2.62$ \\
2.0 & $4.77$ \\
3.0 & $8.45$ \\
4.0 & $11.76$ \\
\bottomrule
\end{tabular}
\end{table}

Typically, modems have a wide range of available modulation and coding (MODCOD) schemes that can be used depending on the SNR link budget, which can be dynamically adjusted depending on the conditions of the satellite link. 

\noindent\fbox{%
  \parbox{\dimexpr\linewidth-2\fboxsep-2\fboxrule-3pt\relax}{
    \textbf{Numerical example no. 4:} For the same link, a coding scheme carrying $0.6$ bit/s/Hz in $1$ kHz would yield 600 bit/s before overhead if it can operate at the calculated SNR. Its theoretical minimum is $10\log_{10}(2^{0.6}-1)\simeq-2.88$ dB, leaving only about 0.57 dB above the ideal threshold. A practical link budget requires the modem threshold at the target error rate.
  }%
}

\subsection{Increasing capacity with multiple beams per satellite and frequency reuse}

Table~\ref{tab:constellations} contrasts reported deployment status with announced plans. Historical shell parameters in~\cite{hypatia} are not necessarily current. Counts refer to different populations (in orbit, operational or launched), as indicated, and cannot be compared as a common active-satellite inventory.

\begin{table}[htbp]
\centering\small
\caption{Constellation status reported by operators, as of 18 November 2026. Counts are rounded or qualified as reported.}
\label{tab:constellations}
\begin{tabular}{p{2.5cm}p{3.6cm}p{7.2cm}}
\toprule
System & Reported scale/status & Deployment and service context \\
\midrule
Starlink & $>$11,000 in orbit (Oct. 2026) & Operational broadband; direct-to-cell is a separate service~\cite{starlink_update,starlink_mobile}. \\
Eutelsat OneWeb & $>$600 in orbit; 12 planes at 1,200 km & Operational business and mobility broadband; 340 replacement satellites ordered in Jan. 2026~\cite{oneweb2026}. \\
Amazon Leo & 396 deployed on 14 successful missions & More than 3,000 planned. The 15 November Vulcan mission failed to reach orbit~\cite{amazon_deployed2026,ula_leo2026,amazon2026}. \\
Telesat Lightspeed & 225 planned & Expansion announced in Aug. 2026; global service expected in Q1 2028, subject to programme execution~\cite{telesat2026}. \\
IRIS$^2$ & 348 planned (330 LEO, 18 MEO) & European multi-orbit programme, not yet an operational 348-satellite constellation~\cite{iris2026}. \\
\bottomrule
\end{tabular}
\end{table}

\textbf{Numerical example no. 5:} For $N$ evenly spaced satellites in a circular orbital plane of altitude $h$, each adjacent pair subtends an angle of $2\pi/N$ radians at the centre of Earth. Multiplying by the orbital radius $R_E+h$ gives an arc of $2\pi(R_E+h)/N$. Ground coverage depends on elevation masks, Earth curvature, antenna beams and link budget. For example, with $N=22$, $h=550$ km and $R_E=6371$ km the separation is approximately 1977 km along the orbital path. This is a hypothetical geometry, not an assertion about a current Starlink shell.

Each LEO satellite is often equipped with multiple antenna beams pointing at different regions (or cells) in its footprint area, and allowing frequency reuse like in mobile networks. This can be achieved in multiple ways, a typical one is by using phased array antennas with high directivity. Indeed, High or Very-High Throughput Satellites (HTS/VHTS) represent an evolution of satellites towards higher capacity through more spot beams and higher frequency reuse. These, applied in LEO orbit constellations, can further provide high bandwidth and reduced latency to enable Mobile Broadband (MBB) and Machine-Type Communications (MTC) in places where both fibre and 5G connectivity has limitations (deep rural areas, sea-side, mountains, etc). V/HTS can be identified by two key technological features~\cite{guan_review}:
\begin{itemize}
    \item The use of multiple spot beams (tens, even hundreds) of narrow beams covering a small geographical area cells, as shown in Fig.~\ref{fig:arch}. 
    \item The frequency reuse of allocated bandwidth in non-adjacent beams/cells, thus higher throughput of the satellite.
\end{itemize}

Partitioning a region into cells served by individual spot beams allows non-adjacent cells to reuse the same frequencies, provided that the signal-to-interference-plus-noise ratio (SINR) remains high enough for the chosen waveform. The 3GPP simulation assumptions illustrate frequency and polarisation reuse~\cite[Table~6.1.1.1-5]{tr38821}. Dividing the usable spectrum equally among $N_c$ frequency groups and summing the rate over all simultaneous beams and polarisations gives the following illustrative estimate:
\begin{equation}\label{eq:reuse_capacity}
    R_{tot} = \beta B_w \left( \frac{N_p N_b}{N_c} \right)  \cdot (1-\eta_{guard})
\end{equation}
Here $\beta$ is the same net spectral efficiency for each active beam, $B_w$ is the total available bandwidth per polarisation across all colours, and beams are equally loaded. This is a derived accounting estimate, not a measured bound; simultaneous beams, interference, payload power and gateways may limit throughput. $N_p$ stands for the number of polarizations (1 or 2), $N_b$ is the number of spot beams (several tens, even hundreds for VHTS), $N_c$ is the number of colors or frequencies (3, 4 or 6 typically), and $\eta_{guard}$ is the fraction of $B_w$ reserved for guard bands.

\noindent\fbox{%
  \parbox{\dimexpr\linewidth-2\fboxsep-2\fboxrule-3pt\relax}{
    \textbf{Numerical example no. 6:}  Consider a satellite operating in the Ku-band with 1.5~GHz of spectrum per polarisation and a spectral efficiency of 2~bit/s/Hz. With two polarisations per beam, seven frequency groups, 60 simultaneous spot beams and a 10\% guard-band fraction, the simplified estimate is:
    \[R_{tot} = 2 \cdot 1.5\cdot 10^9 \frac{2\cdot 60}{7} \cdot 0.9 \simeq 46.29~\mathrm{Gb/s}.\]
    }
}

Ku- and Ka-band systems can exploit wider spectrum and spatial reuse, subject to interference, rain attenuation, gateway capacity and licensing. Aggregate satellite capacity, per-beam bandwidth and achievable user throughput are different quantities and should not be interchanged.

Multiple spot beams also create payload, power, processing and cost trade-offs. Historical empirical cost curves should not be extrapolated to current satellite generations without validating their assumptions~\cite{guan_review}.

In general, when using multiple beams, each single beam may interfere with adjacent ones in the frequency of operation. The final Signal to Interference and Noise Ratio (SINR) is obtained by adding uncorrelated noise and interference powers in linear units~\cite[eq.~(6.1.3.1-1), expressed here in linear units]{tr38821}:
\begin{equation}\label{eq:sinr}
    SINR = \frac{S}{N+I} = \frac{1}{\frac{1}{SNR}+\frac{1}{SIR}}
\end{equation}
Here $S$, $N$ and $I$ are the received signal, noise and aggregate interference powers over the same bandwidth. The second equality uses the linear ratios $SNR=S/N$ and $SIR=S/I$.

Spot-beam antennas concentrate power on their intended cells. Interference in neighbouring cells depends on the sidelobe pattern, beam geometry, polarisation and frequency reuse~\cite{guan_review,array_article}.

\noindent\fbox{%
  \parbox{\dimexpr\linewidth-2\fboxsep-2\fboxrule-3pt\relax}{
  \textbf{Numerical example no. 7:} With SNR $=9$ dB and SIR $=6$ dB, the linear ratios are $10^{0.9}\simeq7.943$ and $10^{0.6}\simeq3.981$. Thus $SINR=[1/7.943+1/3.981]^{-1}\simeq2.652$, or $4.24$ dB. If SIR rises to 12 dB with the same SNR, $SINR\simeq5.291$, or $7.24$ dB.
  }%
}

\subsection{Phased Array antennas, radiation pattern and directivity}

To achieve highly directive antennas on board, the use of linear or planar phased arrays of $N$ elements is often considered~\cite{array_article}. As an example, consider an equally spaced linear array (ESLA), where $N$ array elements are separated by some distance $d$, for a total distance of $D=(N-1)d$. Summing the element phasors as a finite geometric series gives the array factor (AF)~\cite{array_article,array_article_toronto}: 
\begin{equation}\label{eq:array_factor}
    AF_{ESLA}(\psi) = e^{j(N-1)\frac{\psi}{2}} \frac{\sin\left(N\frac{\psi}{2}\right)}{\sin\left(\frac{\psi}{2}\right)}
\end{equation}
where $\psi = kd \cos(\theta)$; here $k=2\pi/\lambda$, $d$ is element spacing, and $\theta$ is measured from the array axis. The formula assumes equal amplitudes and zero progressive phase shift; $AF=1$ for a single element.

For a broadside, uniformly excited linear array of $N$ uncoupled isotropic elements spaced exactly $d=\lambda/2$ in free space, the peak array-factor amplitude is $N$. Averaging $|AF|^2$ over the sphere gives $N$, because the cross terms of the phasor sum integrate to zero over $\cos\theta\in[-1,1]$. The peak directivity is therefore~\cite{array_article_toronto}:
\begin{equation}\label{eq:linear_directivity}
    D_{ESLA} = N    
\end{equation}

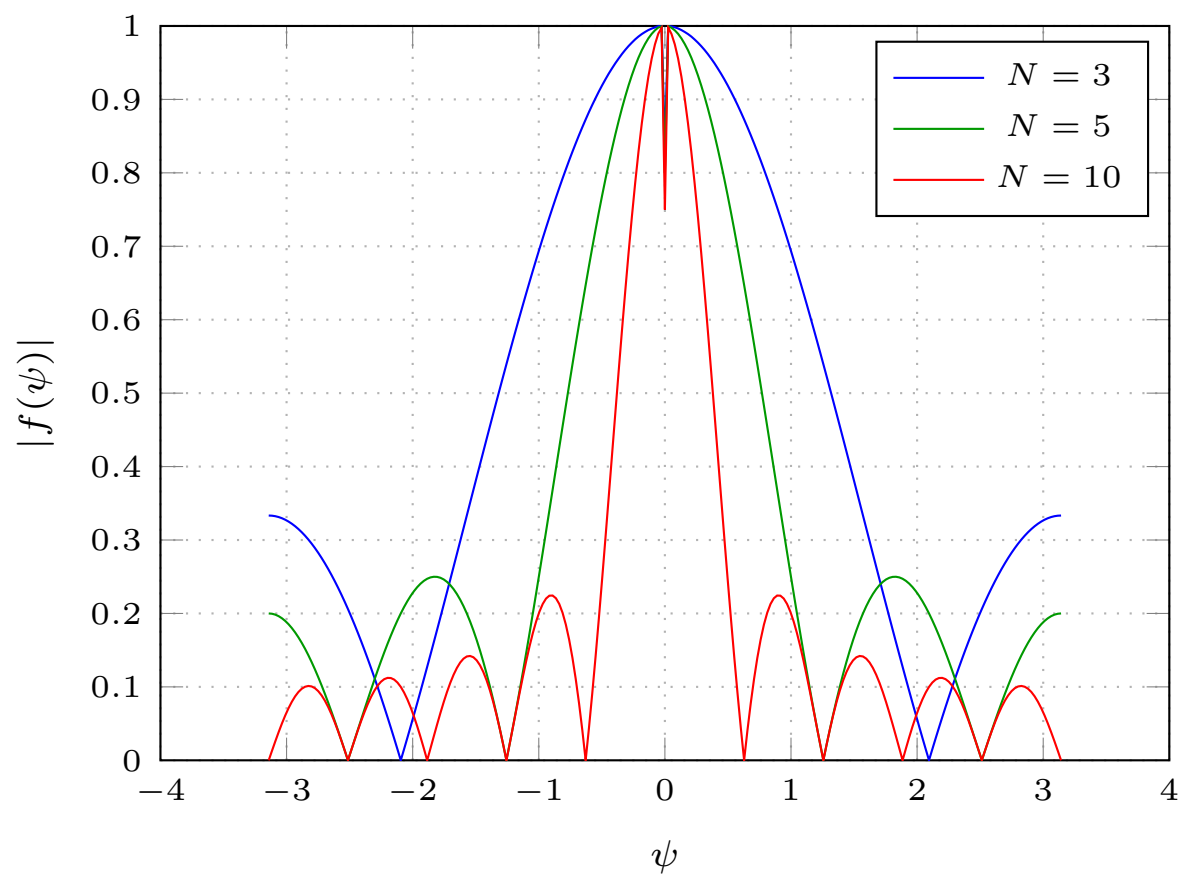
\begin{figure}
    \centering
\begin{tikzpicture}
\begin{axis}[
  width=0.5\textwidth, height=0.39\textwidth,
  xmin=-4, xmax=4, ymin=0, ymax=1,
  xlabel={$\psi$}, ylabel={$|f(\psi)|$},
  xtick={-4,-3,-2,-1,0,1,2,3,4},
  ytick={0,0.1,0.2,0.3,0.4,0.5,0.6,0.7,0.8,0.9,1}, grid=major,
  grid style={dotted,gray!60},
  tick label style={font=\tiny}, label style={font=\scriptsize},
  legend style={at={(0.98,0.98)},anchor=north east,font=\tiny,fill=white},
  samples=241, domain=-3.14159265:3.14159265,
]
\addplot[blue,thin] {abs(x)<0.00001 ? 1 : abs(sin(deg(3*x/2))/(3*sin(deg(x/2))))};
\addlegendentry{$N=3$}
\addplot[green!60!black,thin] {abs(x)<0.00001 ? 1 : abs(sin(deg(5*x/2))/(5*sin(deg(x/2))))};
\addlegendentry{$N=5$}
\addplot[red,thin] {abs(x)<0.00001 ? 1 : abs(sin(deg(10*x/2))/(10*sin(deg(x/2))))};
\addlegendentry{$N=10$}
\end{axis}
\end{tikzpicture}

    \caption{Plots of $|f(\psi)|$ for various $N$~\cite{array_article_toronto}.}
    \label{fig:AFs}
\end{figure}

Hence, disregarding the unit-magnitude phase factor $e^{j(N-1)\psi/2}$ and normalising, we obtain:
\begin{equation}\label{eq:array_factor_normalised}
    f(\psi) = \frac{\sin\left(N\frac{\psi}{2}\right)}{N\sin\left(\frac{\psi}{2}\right)}
\end{equation}
where the broadside value is defined by continuity as $f(0)=1$. The full radiation pattern also depends on the element pattern and environment. Fig.~\ref{fig:AFs} shows examples of AF for different values of $N$.

\begin{figure}
    \centering
\begin{tikzpicture}[scale=1.85,font=\tiny]
\foreach \r in {0.5,1} {\draw[gray, densely dotted] (0,0) circle (\r);}
\draw (0,0) circle (1);
\foreach \a in {0,30,...,330} {\draw[gray,densely dotted] (0,0)--(\a:1); \node at (\a:1.12) {\a};}
\node[fill=white,inner sep=0.4pt] at (70:0.52) {0.5};
\node[fill=white,inner sep=0.4pt] at (78:1.08) {1};
\draw[thick,domain=0:180,samples=241,variable=\t]
 plot ({abs(1+2*cos(180*cos(\t))+2*cos(360*cos(\t)))/5*cos(\t)},
       {abs(1+2*cos(180*cos(\t))+2*cos(360*cos(\t)))/5*sin(\t)});
\end{tikzpicture}
    \caption{Upper-half-plane cut of the normalised array factor for $N=5$ and $d=\frac{\lambda}{2}$~\cite{array_article_toronto}.}
    \label{fig:n5}
\end{figure}
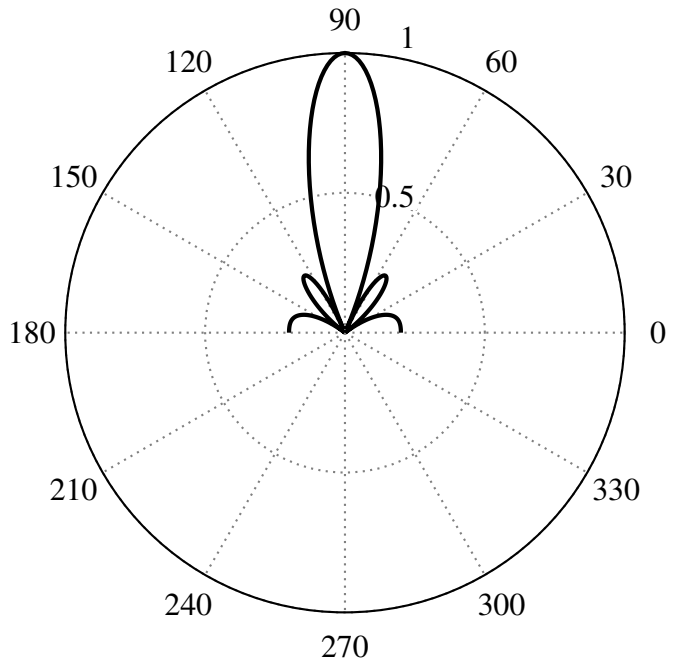

As the number of array elements $N$ increases, the width of the main lobe in the radiation pattern decreases, making antennas more directive (with higher gain and smaller area covered). The sidelobe level depends on the amplitude taper and element pattern; increasing $N$ alone does not generally suppress the first sidelobe of a uniformly weighted array~\cite{array_article}. For example, for $N=5$ and $d=\frac{\lambda}{2}$, the ideal isotropic-element model gives a linear directivity of $D_{ESLA}=5$ and the array factor shows sidelobes (see Fig.~\ref{fig:n5}). The Half-Power Beam Width (HPBW) is the full angular separation between the two directions where power is half its peak ($-3$ dB).

Planar rectangular arrays have a two-dimensional array factor. For a large, uniformly illuminated rectangular aperture, approximate directivity scales with electrical aperture area as $D\simeq 4\pi A/\lambda^2$ under the ideal-aperture model~\cite{continuous_apertures}. The directivity of a finite array depends on element patterns, mutual coupling, taper and back radiation.

The gain of an antenna is the directivity multiplied by its radiation efficiency $k_{ef}$~\cite{array_article}:
\begin{equation}
    \label{eq:gain_directivity}
    G = k_{ef}D\quad \textrm{or}\quad G_{dBi}=D_{dBi}+10\log_{10}(k_{ef})
\end{equation}
Radiation efficiency $k_{ef}=P_{rad}/P_{in}$ is the ratio of radiated power to accepted input power. For $k_{ef}=0.5$, gain is 3.01~dB below directivity. Realised gain also accounts for input mismatch~\cite{array_article}.

For a matched, co-polarised receiving antenna, the effective aperture is related to its linear gain by~\cite{friis1946}: 
\begin{equation}\label{eq:effective_aperture}
    A_e = \frac{\lambda^2}{4\pi} G
\end{equation}
An ideal isotropic antenna has unit gain ($G=1$, or $0$ dBi). 


The peak directivity $D_0$ follows from integrating the radiation intensity $U$ over the sphere~\cite[Sec.~1.C]{planar_array_ref}:
\begin{equation}\label{eq:directivity_integral}
    D_0=\frac{4\pi}{\displaystyle\int_{4\pi} \frac{U(\theta,\phi)}{U_{max}}\,\mathrm{d}\Omega}.
\end{equation}
Here $\mathrm{d}\Omega$ is the solid angle element and $U_{max}$ is the peak radiation intensity. For a large, nearly broadside planar array, one empirical approximation is $D_0\approx 32400/(\theta_x\theta_y)$, with $\theta_x$ and $\theta_y$ the full principal-plane half-power beamwidths in degrees~\cite[eqs.~(16.22) and (16.26)]{planar_array_ref}.


\begin{table}[htbp]
\centering
\caption{Broadside HPBW of uniform half-wavelength arrays, calculated from the normalised array factor in the relevant principal plane~\cite{array_article_toronto,array_article}.}
\begin{tabular}{lcc}\toprule
Array & Model directivity & HPBW \\\midrule
Linear $N=3$ & $3$ (4.77 dBi) & $36.18^\circ$ \\
Linear $N=7$ & $7$ (8.45 dBi) & $14.67^\circ$ \\
Linear $N=11$ & $11$ (10.41 dBi) & $9.27^\circ$ \\
Planar $4\times4$ & Pattern-dependent & $26.32^\circ$ \\
Planar $8\times8$ & Pattern-dependent & $12.80^\circ$ \\
Planar $16\times16$ & Pattern-dependent & $6.36^\circ$ \\
Planar $32\times32$ & Pattern-dependent & $3.17^\circ$ \\\bottomrule
\end{tabular}
\label{tab:array_examples}
\end{table}
The linear directivities assume uncoupled isotropic elements in free space. The table obtains each HPBW by solving $|f(\pi\sin\delta)|^2=1/2$ for the first positive angle $\delta$ from broadside and doubling that angle. For planar arrays this calculation applies in either principal plane under separable uniform weights and isotropic element factors~\cite{array_article,planar_array_ref}. Real gain and directivity require the full 3D pattern and efficiency.

\noindent\fbox{%
  \parbox{\dimexpr\linewidth-2\fboxsep-2\fboxrule-3pt\relax}{
    \textbf{Numerical example no. 8:} For a satellite at 500 km directly above the centre of a cell of radius 50 km, a flat-Earth approximation gives an area of $\pi(50)^2\simeq7854$ km$^2$ and a full subtended angle $2\tan^{-1}(50/500)=0.1993$ rad $=11.42^\circ$. Interpreting this as a desired full HPBW is a design convention, not a sharp coverage boundary. The ideal $8\times8$ array has a $12.80^\circ$ principal-plane HPBW. For its principal-plane cut, Eq.~\ref{eq:array_factor_normalised} gives an array-factor power about $2.36$ dB below the peak at the intended edge, $5.71^\circ$ from boresight. The half-power point lies farther out. Actual coverage depends on the full antenna pattern and link budget.
  }%
}

Beam directivity helps illuminate the target cell while controlling interference. For a uniformly weighted large linear array the first sidelobe approaches approximately $-13$ dB relative to peak power; tapering trades sidelobe suppression against gain and beamwidth~\cite{array_article}.


\section{Recent deployments, capacity and NTN integration}
\label{sec:stin}

\subsection{Broadband constellations and their services}
Starlink operates a large LEO broadband constellation and reports inter-satellite laser links. OneWeb serves enterprise, aviation and maritime users through its ground and space segment; its first-generation constellation is at approximately 1,200 km~\cite{starlink_update,oneweb2026}. Amazon began full-scale Leo deployment in April 2025, renamed the project in November 2025 and launched an enterprise preview. Amazon reported 396 spacecraft deployed over 14 launches before the 15 November Vulcan flight. United Launch Alliance confirmed that the flight carrying the next 40 spacecraft failed to reach orbit~\cite{amazon_deployed2026,amazon2026,ula_leo2026}. Telesat's announced 225-satellite Lightspeed plan and the EU's 348-satellite IRIS$^2$ programme are future systems, not equivalent to operational networks~\cite{telesat2026,iris2026}.

\begin{table}[htbp]
\centering\small
\caption{Examples of service capability and published bandwidth/rate information. Values are reported or advertised by providers; rates are not reserved spectrum or guaranteed performance.}
\label{tab:services}
\begin{tabular}{p{2.5cm}p{4cm}p{6.8cm}}
\toprule
System & Access type/status & Published capacity or spectrum context \\
\midrule
Starlink Mobile & LTE direct-to-cell with partner operators & Messaging and selected apps. The US T-Satellite service includes WhatsApp voice and video on compatible devices~\cite{starlink_mobile,tmobile_apps2025}. \\
Amazon Leo & Dedicated fixed/enterprise terminals; deployment and preview & Nano up to 100 Mb/s DL; Pro up to 400 Mb/s DL; Ultra up to 1 Gb/s DL and 400 Mb/s UL~\cite{amazon_rates}. \\
Eutelsat OneWeb & Dedicated enterprise/mobility terminal & LEO broadband; actual throughput depends on terminal and service agreement~\cite{oneweb2026}. \\
Telesat Lightspeed & Planned enterprise/government system & Initial satellites include 500 MHz Mil-Ka spectrum; not a per-user throughput~\cite{telesat_spectrum}. \\
\bottomrule
\end{tabular}
\end{table}

A MHz spectrum allocation, occupied channel bandwidth, spectral efficiency in bit/s/Hz, per-beam capacity and end-user application throughput measure different things. Sector capacity depends on polarisation, reuse and scheduling, while a terminal's practical rate varies with shared load, service plan, beam geometry, weather, gateway and backhaul. The Amazon values in Table~\ref{tab:services} are advertised peak rates, not independent measurements or a direct-to-phone service. Q/V bands can support feeder links but incur substantial atmospheric attenuation; optical inter-satellite links avoid RF spectrum constraints at the cost of precise pointing and acquisition.

\subsection{Direct-to-device: what is available}
Direct-to-cell connects a compatible handset to a satellite using spectrum and service arrangements with a mobile operator. Starlink uses a satellite-borne LTE base station integrated with partner networks. SpaceX reported successful video-call trials in early 2025. T-Mobile subsequently announced commercial WhatsApp voice and video on its US T-Satellite service in October 2025. SpaceX's product roadmap lists native calling as a future capability. Commercial app support varies with the country, operator, device and sky visibility~\cite{starlink_test,starlink_mobile,tmobile_apps2025}.

\subsection{Integration with 5G and the path to 6G}
3GPP Release 17 introduced normative NR-NTN for 5G broadband and LTE-based IoT-NTN, initially focusing on transparent satellite payloads and GNSS-capable devices. Transparent payloads relay radio signals to a terrestrial base station; regenerative payloads move some processing onboard. Release 18 and 19 extend NTN support, including mobility, coverage and radio-frequency work. Release 20 continues 5G-Advanced NTN work and includes studies on inter-system handover~\cite{3gpp_releases,ntn_review2026,ran113}. Network design must address large and time-varying propagation delay, Doppler, timing advance, satellite and beam handovers, feeder link continuity, location-assisted pre-compensation and mobility across terrestrial and satellite coverage. Authentication, roaming, policy and quality-of-service integration are as important as radio reachability.

A commercial LTE direct-to-cell network is not automatically a deployment of 5G NR-NTN, even when the satellite operator also offers 5G backhaul. The appropriate comparison is the implemented air interface, core-network interfaces, device capability and operator agreement. This distinction also applies to quoted capacity and latency: published peak figures are not guarantees of 5G-like service.

The 6G direction includes more tightly coordinated terrestrial, airborne and space layers, multi-orbit routing, dynamic resource allocation, onboard processing and resilience. These are research and standardisation directions, rather than completed 6G services. 3GPP identifies Release 20 as a 6G study phase and Release 21 as the beginning of normative work; timelines remain subject to change~\cite{3gpp_6g}. ESA's STARGATE project, ongoing in 2026, is developing an end-to-end demonstrator to evaluate candidate 6G NTN waveforms against 5G under representative satellite impairments~\cite{esa_stargate}. Recent work surveys 3GPP NR-NTN releases and models end-to-end performance~\cite{ntn_review2026}, while a 2024 overview frames the transition towards 6G~\cite{sixg_review2024}.

\subsection{Research and deployment challenges}
Important open questions include interference coordination between terrestrial and satellite systems sharing spectrum; robust operation without reliable GNSS; mobility across beams and satellite gateways; energy and uplink constraints of handheld devices; cybersecurity, privacy and resilience; and congestion and fair sharing under highly variable demand. Constellation design also requires collision avoidance, disposal and coordination with astronomy. Independent reproducible measurements across locations, terminal types, loading and weather are needed to distinguish research results and commercial targets from operational performance.

\section{Discussion and future work}
\label{sec:conclusions}

This article has briefly overviewed the use of satellites and HAPs for providing connectivity in those areas where fiber cannot reach, namely deep rural areas, mountains, desert and seaside. The foundations regarding link budget analysis and achievable bitrates are also reviewed, along with an introduction to antenna array systems and cellular designs. The 2026 update distinguishes commercial broadband, direct-to-device service, planned constellations and evolving 3GPP specifications. Research towards 6G remains a developing programme; further work is needed on independently measured performance, interoperability and orbital sustainability.










\section*{Funding} 

The authors acknowledge support from the Spanish national research project \emph{Agents for Networks and Networks for Agents} (ANNA), funded by the Agencia Estatal de Investigaci\'{o}n (\href{https://doi.org/10.13039/501100011033}{AEI}) through its 2025 Knowledge Generation Projects call. The coordinated project comprises grant PID2025-169490OB-C21 at Universidad Polit\'{e}cnica de Madrid and grant PID2025-169490OB-C22 (\emph{ANNA: Networks and Agents}) at Universidad Carlos III de Madrid. The project runs for 36 months, from 1 September 2026 to 31 August 2029.


\end{document}